RESEARCH ARTICLE

# A Forensically Sound Adversary Model for Mobile Devices

Quang Do☯, Ben Martini☯, Kim-Kwang Raymond Choo*

Information Assurance Research Group, University of South Australia, Adelaide, Australia

☯ These authors contributed equally to this work.
* raymond.choo@unisa.edu.au

## Abstract

In this paper, we propose an adversary model to facilitate forensic investigations of mobile devices (e.g. Android, iOS and Windows smartphones) that can be readily adapted to the latest mobile device technologies. This is essential given the ongoing and rapidly changing nature of mobile device technologies. An integral principle and significant constraint upon forensic practitioners is that of forensic soundness. Our adversary model specifically considers and integrates the constraints of forensic soundness on the adversary, in our case, a forensic practitioner. One construction of the adversary model is an evidence collection and analysis methodology for Android devices. Using the methodology with six popular cloud apps, we were successful in extracting various information of forensic interest in both the external and internal storage of the mobile device.

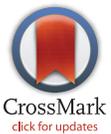

**Data Availability Statement:** All relevant data are within the paper.

**Funding:** The authors have no support or funding to report.

**Competing Interests:** The authors have declared that no competing interests exist.

## Introduction

Human beings have become increasingly dependent on information and communication technologies (ICTs) not just for work and commercial functions, but also for many daily activities. The advent of Web 2.0, since around the mid-2000s, marked the introduction of pervasive interconnectivity, otherwise known as an 'always-on' lifestyle. This has made ICT (e.g. mobile devices and apps) paramount to everyday life, such that any significant technical glitch or outage in internet access or mobile transmission would lead to mass frustration, confusion, and even panic.

Holt and Bossler [1] explained that "[a]s technology increasingly permeates all facets of modern life, there are substantive risks to the safety of digital information and computer networks". For example, the increasing popularity of mobile devices including smart phones (e.g. iOS and Android devices) constitutes an opportunity for cybercriminals.

Technological advances should not, however, be blamed for any increase in criminal exploitation of ICT. Cyber threats and windows of vulnerability evolve over time, partly in response to defensive actions or crime displacement. Therefore, we need to maintain persistent pressure on cybercriminals to safeguard the security and privacy of our data (e.g. stored on mobile devices) even though we may never be able to completely eradicate cybercrime.

To increase the risks of detection and successful prosecution due to the ability to collect evidence from mobile devices, it is important to stay ahead of the race between device (i.e. hardware) and software releases by providers, and software and hardware modifications made by end users to complicate or prevent the collection and analysis of digital evidence.

During investigations of crimes involving mobile devices, there is usually some accumulation or retention of data on the device that will need to be identified, preserved, analyzed and presented in a court of law—a process known as digital or mobile forensics (also known as cyber forensics) [2].

Potential evidential data that could be recovered from mobile devices includes login credentials for cloud storage and other online accounts, and metadata such as when and where a photo or video was taken (i.e. geolocation) using the mobile device or stored in the cloud.

Studies of digital forensic toolkits currently available to law enforcement agencies reveal significant challenges and complications in extracting evidence from newer devices, with attempts to extract information from various mobile devices using a range of mobile forensic tools producing differing results [3, 4].

There is little doubt that governments will continue to be under pressure to deliver more with less. This will apply to the provision of law enforcement services no less than other government services. Therefore, to keep pace with this growth and the changing face of criminal activity, particularly to ensure that evidential data can be forensically recovered from mobile devices, we propose a forensic adversary model for mobile device data collection and analysis. Forensic practitioners, like adversaries, are subject to certain constraints. In the case of a forensic practitioner, these constraints relate to "forensic soundness". These constraints help to ensure that evidence remains valid and suitable for presentation to a court of law and that it remains untainted. We integrate the constraints of forensic soundness into our adversary model, resulting in a forensically sound adversary model.

The remainder of the paper is organized as follows. The next section discusses related work. Section 3 outlines our research motivations. We then present our mobile forensic adversary model and a construction of this model, the Android evidence collection and analysis methodology (see also [5]), in Section 4. We discuss the utility of our construction in Section 5. The last section concludes this paper and outlines future work.

## Related Work

### 2.1 Adversary Models

Adversary models have long been an important part of designing and validating cryptographic and secure systems (e.g. used to prove or disprove the properties of a cryptographic protocol). An adversary generally has capabilities, such as the ability to listen to all messages transmitted by a target. One of the most recognized and utilized adversary models is, perhaps, the Dolev-Yao model [6]. An adversary in this particular model has capabilities that verge on omnipotence, including the ability to intercept messages from a target, modify these messages and even initiate a connection with the target. Similarly, Bellare and Rogaway's adversary model [7] provided the adversary with comparable capabilities. The adversary has the capacity to control all communications between the interacting parties. All messages produced by the parties are observable and modifiable by the adversary. The adversary is even capable of controlling the delays between messages, replaying messages, employing multiple sessions and corrupting an honest user in the system. Some may criticize that both adversary models are too strong, and hence, unrealistic.

Adversary models have also been used in network security to replicate and simulate attacks on communications protocols. Syverson, et al. [8] presented an adversary model for networks utilizing onion routing. The capabilities of this model include the ability to: observe (sniff) a connection, delay or corrupt a connection, initiate or destroy a connection and modify a connection's traffic at will. This model differs somewhat from those designed for cryptographic protocols. The adversary's main aim is to compromise connections between parties instead of manipulating the data of the actual traffic itself. Similarly to the cryptographic-based adversary models, this model is somewhat specific to the field of network security.

In the field of privacy preservation, adversary models have been used to model privacy invasions. Heiber and Marron [9] considered the privacy of users in context-aware systems and presented a framework for the modelling of privacy within these systems. In order to demonstrate their model, the authors utilized an adversary model capable of over-hearing all wireless transmissions in a network and also having full access to the context-aware system's location service. The specificity of this model makes it difficult to use outside of this research.

Within the field of Android security, a number of researchers have also utilized adversary models within their work. For example, both Wu, et al. [10] and Zhou, et al. [11] utilize adversary models that involve the adversary simulating a malicious third-party app installed on an Android device. The adversary model of Zhou, et al. [11] can only request non-sensitive permissions and the adversary model of is unable to request for any permissions. Such specific adversary models cannot be used to generalize or model other types of security attackers.

Other research that makes use of adversaries often informally defines an adversary or an adversary's capabilities, or refers to the adversary in passing. The use of loosely defined adversaries (e.g. a "strong adversary") and/or adversary capabilities means that the research lacks a concrete adversary which would be useful in other research environments. Bugiel, Heuser and Sadeghi [12] presented a security architecture for access control on Android devices. In doing so, the authors provided an adversary model that exhibited a "strong" adversary able to "access sensitive data" and "launch software attacks". Although their research does not focus on the adversary model, the formalization concerning the capabilities of their proposed adversary model is somewhat lacking.

As described above and shown in Table 1, there are many different use cases for adversary models (with varying levels of competence). These different use cases also prompt researchers to design their own adversary models that may be incompatible with other research both inside and outside of their field of study. For example, Syverson, et al. [8] focused on the security of onion routing networks which are designed to provide anonymity. Cryptographic protocols, on the other hand, are intended to provide secure communications between parties—the identity of each party would be known before communication had even commenced. Further differences arise when comparing a cryptographic or networking adversary to a smartphone adversary. Take, for example, a user with a smartphone which is connected to an onion routing network who wishes to perform a secure transaction with another party. A smartphone adversary typically wishes to obtain sensitive user data (e.g. contact phone numbers, location information and login details), message or call premium numbers or otherwise obtain some data of (monetary) value whilst the onion routing adversary seeks to determine the identity of the user. The cryptographic adversary would like to decrypt the secure communications between the smartphone and the other party. Table 1 demonstrates the differences in adversary models with regards to the ways they are defined, including their goals and assumptions (and the levels of detail provided).

### 2.2 Android Data Collection

There have been a number of data collection methodologies designed specifically for Android devices. These are a combination of techniques (methods or approaches), often with an underlying process (model or framework) in order to bring them together.

Table 1. A comparison of adversary models.

| | Adversary Capabilities Fully Defined? | Adversary Goals | Adversary Assumptions | Adversary Model Limitations |
|---|---|---|---|---|
| Dolev and Yao [6] (Cryptographic protocols) | Yes | The adversary seeks to obtain the plaintext for encrypted messages on the network. | The adversary can read any message on the network, and can initiate a two-way connection with any other user. | Very powerful, making it difficult to model weaker adversaries. |
| Bellare and Rogaway [7] (Cryptographic protocols) | Yes | Similarly to the Dolev-Yao model, the adversary aims to obtain the long-term secret key and/or the ephemeral / session key. | The adversary has similar capabilities to the Dolev-Yao model, including the ability to delay or replay messages. | Very powerful, but also flexible due to the notion of "queries" (i.e. adversary capabilities). |
| Syverson, et al. [8] (Network security) | No | The aim of the adversary is to determine the identity of (i.e. de-anonymize) the traffic flowing through its compromised nodes. | The adversary is assumed to have control of one or more nodes (known as Core Onion Routers). | Adversary capabilities are not formally defined. Furthermore, this model is specific to onion routing (e.g. compromised nodes), and cannot be generalizable. |
| Heiber and Marron [9] (Network security) | Yes | The adversary's goal is to collect personal or private data (e.g. location information) from a network. | The adversary is designed such that it is only able to view data (i.e. it cannot modify it). | A weak adversary that is formally defined using inference rules. |
| Wu, et al. [10] (Android security) | No | The adversary aims to steal money from the user, gather sensitive information or destroy data on the device. | The adversary cannot initially request any "sensitive" permissions. | The adversary is defined very loosely. This means it cannot be used for formal proofs. |
| Zhou, et al. [11] (Android security) | No | The adversary aims to accurately determine the locations and behaviors of the user. | The adversary cannot request *any* permissions initially. | This adversary is also very loosely defined. |
| Bugiel, Heuser and Sadeghi [12] (Android security) | No | The adversary's goals are to obtain sensitive data and compromise apps on the device (both system and third-party) | The adversary is assumed to be "strong". | This adversary lacks meaningful detail. |

doi:10.1371/journal.pone.0138449.t001

Most notable of the current methodologies proposed is the general collection methodology for Android devices presented by Vidas, Zhang and Christin [13]. It comprises a process for obtaining data from an Android device that relies on a bootable image which is flashed onto the Android device. The images differ based on the phone model and are required to be flashed over the original recovery image on the Android device. This custom recovery mode is able to extract all data from the NAND and any attached SD cards of the device without requiring root access. One major problem with this approach is that newer phones (released after the publication of [13] in 2011) have locked bootloaders, generally requiring the device be wiped in order to unlock them (if indeed there are ways to unlock the bootloader at all).

Lessard and Kessler [14] presented a process for obtaining all the data from an Android NAND (Negated AND) flash. One technique proposed allowed for collecting a bit-for-bit copy of the NAND flash itself in order to perform data recovery on deleted items. Their process requires rooting the device in order to perform a "dd" image of the relevant partitions and storing these on an external SD card attached to the device, and then analyzing these dumps for potential evidence. One issue with this technique is that it requires rooting the device and the disadvantages to the forensic process that it brings in order to create these images and the technique also requires the device to have a microSD card slot. Many popular Android devices, such as the Nexus 5, Samsung Galaxy S6 and LG G2, do not have the ability to accept SD cards, rendering this process infeasible on these devices.

Chen, Yang and Liu [15] suggested that physical data acquisition on an Android device can be performed without modifying the Android device in almost any manner and presented a process to do so. The authors use the Android device's recovery mode in order to run their own "update.zip" (generally used by carriers and phone manufacturers to issue Android OS updates) package which is stored on an external SD card. As the "update.zip" package runs as a root process, it is able to perform a physical data acquisition of the Android device and dump it onto the SD card. Two major problems arise with this suggested method. Firstly and most importantly, this method relies on the Android device having already have flashed a third-party recovery mode on the recovery partition. This is unlikely to be the case as the user would have to unlock the bootloader of the device and flash this software manually. The reason a third-party recovery mode software is required is that first-party recovery modes generally have signature checking which is used to make sure the "update.zip" files were packaged by the appropriate authority (carriers or phone manufacturers). Such an "update.zip" package designed by the authors would refuse to run on first-party recovery modes. In addition to requiring a third-party recovery mode, this method also requires the device have an SD card slot.

A methodology for collecting data from Android devices was presented by Votipka, Vidas and Christin [16]. They suggested that a custom recovery image could be built which can be easily ported to many Android devices. This image contains forensic tools that can extract all the required data in a forensically sound manner from the Android device. In order to bypass a locked bootloader, the authors rely on the forensic practitioner having the ability to obtain the manufacturer key used to sign legitimate bootloader images. In most cases, device manufacturers are unlikely to be in the same jurisdiction as the forensic practitioner and may not be legally obligated to provide the key. This makes it highly infeasible that a forensic practitioner would be able to obtain the key. Furthermore, flashing this custom recovery image overwrites the original first-party recovery mode on the device, reducing the forensic soundness of this process.

Similarly, Son, et al. [17] proposed a data acquisition process for Android that relied on the use of a custom recovery mode image that was flashed onto the boot partition. Flashing the boot partition with a custom recovery mode image will render the device unable to boot into the original Android OS. The authors state that the boot image must be flashed with the original boot image of the device, which may be difficult (and not forensically sound depending on the source) to obtain. They also designed an application that extracts user data stored within the Android device via ADB after the custom recovery mode image was installed. Although their process should be capable of collecting most of the data from the Android device's NAND flash, the authors focus on obtaining user and app data.

Finally, a crucial stage that the majority of contemporary data collection methodologies currently lack is the analysis of the collected data. Without accurate analysis, the collected data cannot be utilized to its full extent. In addition, without an adequate analysis of the collected evidence, the forensic practitioner may omit critical evidential data.

## Research Motivations

As discussed in the previous section, adversary models exist for a range of fields in the security and cryptography disciplines. However, adversary models have not yet been adopted in digital forensics. We hope to contribute to filling this gap in this paper as we believe that the role of the traditional security 'adversary' represented in existing models and a forensic practitioner conducting investigations have significant parallels.

Generally, adversary models are designed for use in a particular field, each of which has different specific requirements and capabilities. This holds true for digital forensic adversary models, where an overriding principle must be maintained, that is, the principle of ensuring that evidence sources remain unmodified wherever possible [2]. Where a change to an evidence source is required to collect evidence, these changes must be discrete (i.e. the practitioner must be able to report upon the exact actions they performed and the transient and persistent effect(s) this had on the device, preferably at the bit level for persistent changes) and kept to an absolute minimum [2]. This requirement significantly constrains the ability for forensic practitioners to utilize a range of invasive attack techniques (particularly attacks developed by unknown individuals, with only compiled code released) that would be available to traditional adversaries.

For example, when collecting evidence from secure devices, practitioners often need to bypass operating system (OS) or hardware security features (e.g. app / process sandboxing). In the case of smartphones, this is achieved via 'rooting' or 'jailbreaking' the device. These processes generally rely upon security flaws to escalate the privileges of the attack code, and in turn, the device's user. However, in most cases, the source code underlying the privilege escalation procedure is not released. This makes it particularly difficult for a forensic practitioner to report on the discrete operations that are being undertaken on the device. In practice, this could result in the destruction or modification of evidence on the device, either intentionally or inadvertently, by the privilege escalation procedure.

In contrast, while many adversary models constrain the adversary by requiring that attacks be executed remotely (e.g. via a network link), such restrictions do not apply to forensic practitioners. In most cases, forensic practitioners are able to 'seize' the physical device being analyzed and, as such, have access to capabilities that require physical access/control of the device. This is a reasonable and necessary ability for forensic practitioners in most cases, even in circumstances where the only apparently feasible option is remote collection of evidence [18]. This is apparent, for example, in the case of cloud computing evidential data, which is commonly stored on remote servers outside of the jurisdiction of the investigating law enforcement agency (LEA). However, despite the primary evidence source being remotely located, evidential data generally exists on the client devices (e.g. laptops, tablets and smartphones) that are used to connect to the remote cloud service. These client devices are generally located within the jurisdiction of the investigating LEA and within the possession or control of the individual being investigated (drawing a physical link between the electronic evidence and the suspect [19]).

The contribution of an adversary model in the field of digital forensics allows for a more formal approach to the development of forensic methodologies. The model can be used as a basis for ensuring that a practitioner can conduct an investigation using only the forensically sound capabilities available to them (e.g. mapping adversary capabilities in combination with an existing forensic model to develop a forensically sound methodology).

## Mobile Forensic Adversary Model

### 4.1 The Proposed Model

As described in Section 3, one of the primary constraints on the forensic adversary is in ensuring that all actions on the evidence item are forensically sound. As such, it should be noted that all of the above capabilities are discrete (i.e. the practitioner is aware of the precise effect they will have on the device being analyzed) and documented as part of standard forensic reporting procedures. Based on the definitions of McKemmish [20] and Casey [21], we formalize the definition of forensic soundness in Definition 1.

**Definition 1** (Forensic Soundness) A process is forensically sound if it satisfies all of the following key criteria described by McKemmish [20] and Casey [21]:

1. *Meaning*: Meaning refers to the fact that evidence that is collected as part of digital forensic investigations must retain its original meaning and interpretation. However, it should be noted that "[i]mposing a paradigm of "preserve everything but change nothing" is impractical and doing so can create undue doubt in the results of a digital evidence analysis, with questions that have no relation to the merits of the conclusions" [21].

2. *Errors*: Errors refer to the need to identify the existence of errors, when they occur, and, explain and justify that their existence has not affected the core validity of the evidence. Hashes are a common method of detecting errors during forensic collection processes.

3. *Transparency and Trustworthiness*: Transparency and trustworthiness highlight the need for independent oversight into the forensic processes used. For example, a court may wish to validate the integrity of the process.

4. *Experience*: Experience refers to the need for individuals undertaking forensic investigations to have sufficient experience, such that their findings can be relied upon.

The forensic soundness criteria (Definition 1) are considered in the selection of available capabilities, outlined below, and specific constraints raised by these criteria are noted as part of our capability descriptions.

Following the approaches of Dolev and Yao [6] and Bellare and Rogaway [7], we introduce an adversary model for mobile forensic data collection and analysis. In this model, there exists an adversary with physical access to a mobile device and capabilities to exploit device security vulnerabilities. The adversary aims to obtain confidential data from a target device within the constraints of forensic soundness (Definition 1). These capabilities are described below:

1. **Corrupt** (*Target device*) allows the adversary to take over the Target device. Such a capability, typically used to capture insider attacks in cryptographic protocols and network security, allows the adversary to learn the complete internal state of the device. Under a strict implementation of the forensic soundness criteria, this capability would be constrained to functions that do not introduce *errors* (e.g. modification of evidential data sources) or result in a loss of *transparency and trustworthiness* (e.g. the execution of unknown operation sequences on the Target device).

2. **Delete** (*Target device*, *Target data*) enables the adversary to delete data (*Target data*) stored on the *Target device*. Under a strict implementation of the forensic soundness criteria, this capability would be significantly constrained to functions that do not result in the introduction of *errors* or a loss of *meaning* (e.g. deletion operations on evidential data sources).

3. **Encrypt/Decrypt** (*Target device*, *Target data*, *Key*) allows the adversary to either encrypt or decrypt a message (*Target data*) from the *Target device* with the *Key*. For the Decrypt function, the decryption key is obtained using another capability (e.g. Forensic Examination). It may not necessarily be feasible to obtain the decryption key for all encrypted items (e.g. the keys may be stored remotely). A strict implementation of the forensic soundness criteria would not significantly affect this capability as it operates on data that has been collected (e.g. Forensic Copy), rather than evidence source data.

4. **Exploit** (*Target device*, *Entry-point*) allows the adversary to exploit the *Target device* using a known exploit. Under a strict implementation of the forensic soundness criteria, this capability would be constrained to functions that do not introduce *errors* or result in a loss of *transparency and trustworthiness*.

5. **Forensic Copy** (*Target device*) allows the adversary to make a logical or physical forensic copy of the *Target device*. This capability intrinsically meets the criteria for forensic soundness. The capability would be designed to ensure that *errors* are avoided and identified where they are encountered as part of the copy procedure.

...

6. **`Forensic Examination`** (*Target device*, *Target data*, *Method*) allows the adversary to undertake a forensic examination of *Target data* on the *Target device* using a specific *Method* (e.g. keyword searches). The operation of the capability would be subject to the forensic soundness criteria, particularly the *meaning* criteria, in ensuring that the true meaning of any data examined is maintained. This is fundamentally related to the *experience* of the forensic practitioner.

7. **`Inject`** (*Target device*, *Entry-point*, *Message*) allows the adversary to inject or infiltrate a *Message* (i.e. binary data such as code) onto the *Target device* via an entry-point (e.g. infiltrated app). For example, an adversary could add code into the Keyboard app (*Entry-point*) on the *Target device* to facilitate other attacks (e.g. `Transmit`) [22]. Under a strict implementation of the forensic soundness criteria, this capability would be constrained to functions that do not introduce *errors* or result in a loss of *transparency and trustworthiness* or *meaning*.

8. **`Listen`** (*Target device*) allows the adversary to passively monitor a communication channel on the *Target device*. This capability must be utilized with consideration for the requirement to obtain data with a known *meaning*, ensure that *errors* in collection are avoided and identified where unavoidable, and that the process for interception is subject to *transparent* review.

9. **`Modify`** (*Target device*, *Target data*, *Message*) allows the adversary to modify an existing message (*Target data*) on a *Target device*, replacing it with a new message (*Message*). An example would be an adversary replacing configuration or modifying app execution on a device. Under a strict implementation of the forensic soundness criteria, this capability would be constrained to functions that do not introduce *errors* or result in a loss of *transparency and trustworthiness* or *meaning*.

10. **`Transmit`** (*Target device*, *Message*) permits the adversary to transmit and/or exfiltrate a *Message* (i.e. binary data such as code and SMS) from the *Target device*. As with the **`Listen`** capability, this capability must be utilized with consideration for the requirement to transmit data with a foreknown effect (i.e. *meaning*), ensure that *errors* in transmission are avoided and identified where unavoidable, and that the process for transmission is subject to *transparent* review.

Based on the definition of forensic soundness and the proposed adversary capabilities, we introduce two levels of forensic soundness: strict and standard. A forensic practitioner requiring a strict level of forensic soundness can only utilize constrained adversary capabilities (with the constraints outlined above). A strict adherence to forensic soundness should be the initial stage in any investigation in order to reduce modification of evidence sources. The less stringent standard level of forensic soundness could be utilized if a strict investigation does not provide sufficient evidence. It is at this level that forensic practitioners should use the capabilities to their full potential to obtain evidential data, while meeting the constraints of forensic soundness (Definition 1).

### 4.2 A Construction

The following section provides a brief outline of our Android evidence collection and analysis methodology [5]–see Fig 1. The methodology is constructed using the mobile forensic adversary model (outlined in Section 4.1).

After identification and preservation of the mobile device has been completed, evidential data collection must be undertaken. Collection commences with the configuration of the

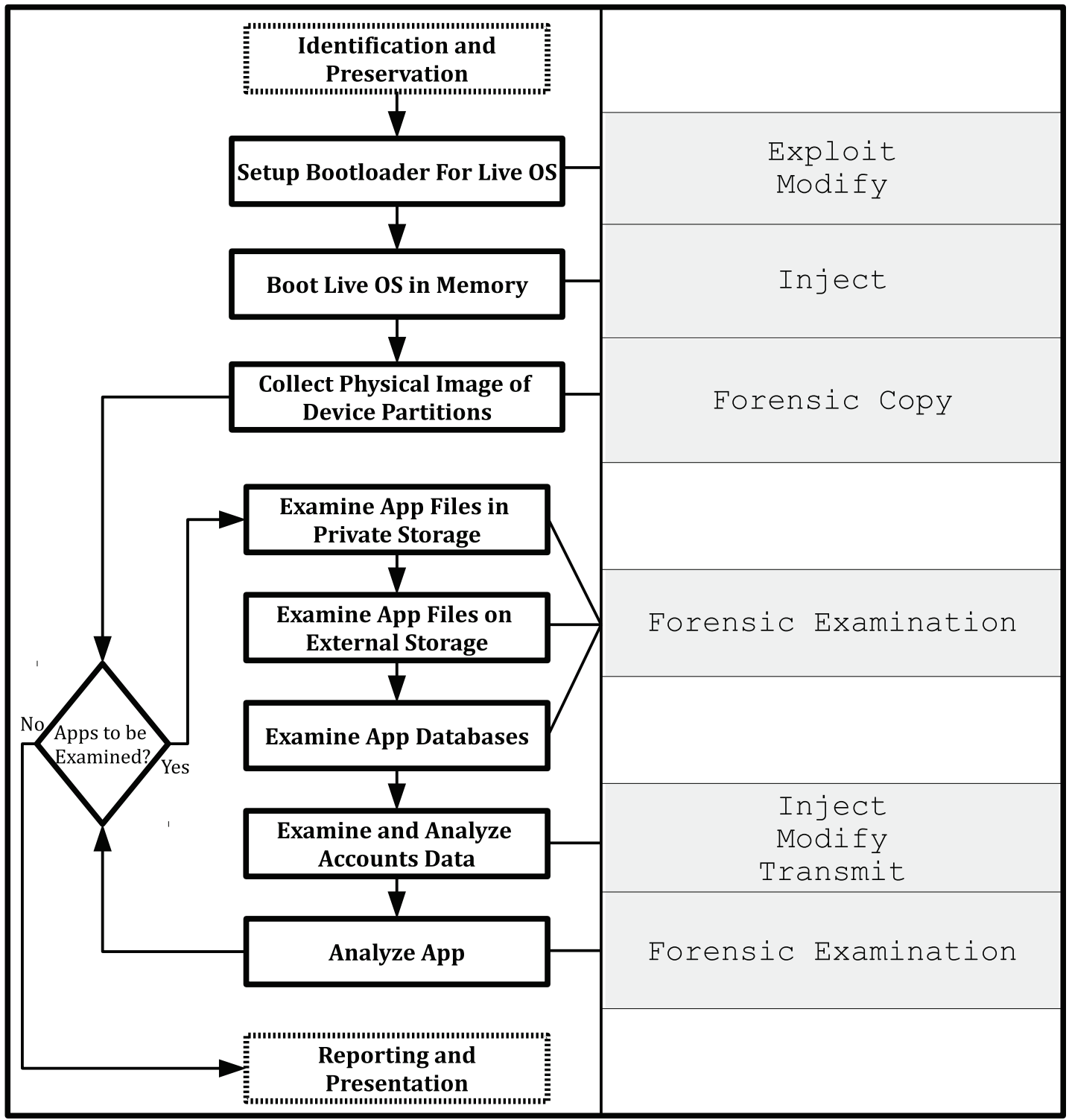

Fig 1. Android evidence collection and analysis methodology (adapted from [5]).

doi:10.1371/journal.pone.0138449.g001

A Forensically Sound Adversary Model for Mobile DevicesTable 2. A mapping of methodology stages to adversary capabilities, constraints and adherence to forensic soundness constraints.

| Methodology Stage [5] | Adversary Capabilities Utilized | Key Forensic Soundness Constraints | Adherence to Forensic Soundness Constraints |
|---|---|---|---|
| Setup Bootloader for Live OS | `Exploit` and `Modify` | *Errors*, *Transparency and Trustworthiness*, and *Meaning* | Implementation dependent for *Errors*, adheres for *Transparency and Trustworthiness*, and *Meaning* |
| Boot Live OS in Memory | `Inject` | *Errors*, *Transparency and Trustworthiness*, and *Meaning* | Adheres to all applicable forensic soundness constraints |
| Collect Physical Image of Device Partitions | `Forensic Copy` | N/A | N/A |
| Examine App Files in Private Storage | `Forensic Examination` | *Meaning* | *Experience* dependent |
| Examine App Files on External Storage | `Forensic Examination` | *Meaning* | *Experience* dependent |
| Examine App Databases | `Forensic Examination` | *Meaning* | *Experience* dependent |
| Examine and Analyze Accounts Data | `Inject`, `Modify` and `Transmit` | *Errors*, *Transparency and Trustworthiness*, and *Meaning* | Adheres to all applicable forensic soundness constraints |
| Analyze App | `Forensic Examination` | *Meaning* | *Experience* dependent |

doi:10.1371/journal.pone.0138449.t002

bootloader, which must be setup to allow the booting of a live OS. This recovery image is booted and then used to collect the physical image(s) of the device partition(s).

Using the images obtained as part of the initial collection stages, we then undertake examination and analysis procedures of a forensic copy of the data stored on the device. Once a practitioner has determined the apps that they wish to investigate, they begin by examining the files stored by the app in its private storage, followed by any files stored in the device's external storage (e.g. SD card). During this examination, it is common to find databases which are examined as part of the next stage. Finally, if the app being examined has an online component (e.g. cloud sync), the account data stored on the device is examined and analyzed to obtain any relevant credentials. If the practitioner has any doubts as to the *meaning* of the data obtained during the examination, they will then undertake an analysis of the app which involves examining the app memory heap in a virtual environment and decompilation of the app code to gain an understanding of the code's operation. Once all relevant data has been obtained and its meaning and provenience (i.e. the source of the evidential data) is fully understood, the practitioner will report and present their findings. The practitioner may also appear as an expert witness in a court of law.

With regards to forensic soundness, the Android evidence collection and analysis methodology should meet the strict level of forensic soundness as defined in Section 4.1, dependent on the implementation used and the *experience* of the forensic practitioner. Even though this methodology requires modifications to be made to the evidence source, all modifications should be fully known by the practitioner and therefore follow the strict adversary capability constraints (see Table 2).

The third and fourth columns of this table are especially pertinent as they provide the specific forensic soundness constraints for a stage and whether the stage adheres to the constraints, respectively. For example, the modification and exploitation of the evidence source to allow the bootloader to accept the forensic practitioner's custom live OS has three constraints: the process must not introduce *errors*, it must be clear (i.e. *transparent and trustworthy*) what modifications are being made and it must not affect the *meaning* of the evidence. Depending on how the device is modified, *errors* may be introduced (e.g. by flashing boot partitions). Ideally, no *errors* should be introduced if the device has been successfully exploited. *Transparency and*

PLOS ONE | DOI:10.1371/journal.pone.0138449  September 22, 2015  10 / 15

*trustworthiness*, and *meaning* of the evidence are not affected by this stage. Further along the methodology are the three stages that relate to examining data on the evidence source's external and internal (private) storage. The *meaning* of the evidence obtained is directly linked to the forensic practitioner's *experience* in dealing with similar devices and knowledge of what constitutes evidential data.

## Discussion

In order to verify the Mobile Forensic Adversary Model when applied to the Android evidence collection and analysis methodology [5], we analyzed a Google Nexus 4 mobile device (see sections 5.1–5.3) and six popular cloud apps, namely, Dropbox, Box, OneDrive, OneNote, Evernote and ownCloud [23] (see sections 5.4–5.8). The relationship between our findings and the adversary's capabilities and forensic soundness constraints is presented in the following subsections.

### 5.1 Setup Bootloader for Live OS

After the identification and preservation stages have been completed, the first step is to `Exploit` the security features in the device bootloader with a view to booting an alternative, forensically sound, operating system on the mobile device. This `Exploit` generally requires the modification (i.e. use of the `Modify` capability) of low level parameters within the device flash, which are configured by default to prevent the execution of unsigned operating systems. Constraints on the `Modify` and `Exploit` capabilities must be adhered to as part of this process, for example, ensuring that *errors* in the evidence source are not inadvertently introduced during the modification of the bootloader configuration parameters. The process used must be subject to *transparent and trustworthy* verification, which limits the capacity for forensic practitioners to utilize existing exploits where the detailed operation of the exploit is unknown.

### 5.2 Boot Live OS in Memory

Once the bootloader has been appropriately configured, the `Inject` capability is used to execute the OS code for the forensically sound operating system in device memory. Similar forensic soundness constraints apply to this stage as those for setting up the bootloader. For example, the operating system used must be *transparently* verifiable, not introduce unavoidable *errors* and be designed by a practitioner with sufficient *experience*.

### 5.3 Collect Physical Image of Device Partition

Using the operating system that has been injected into memory, we are now able to utilize our `Forensic Copy` capability to collect a physical image of the device's flash storage. Hashes (such as MD5 and SHA1) are utilized to verify the forensic soundness of the copy operation and to ensure that *errors* are not introduced in the transmission. Stages that utilize the `Forensic Examination` capability will use the output of the `Forensic Copy` capability.

### 5.4 Examine App Files in Private Storage

We utilized the `Forensic Examination` capability to examine the files stored within the apps' private directories. We were successful in locating a number of notable artefacts, including personally identifiable information (PII), authentication tokens, encryption keys and cached data. In terms of PII, we located usernames, email addresses and geolocation information, along with any PII stored in the files cached by the apps. Authentication tokens were a particularly interesting finding as, combined with the capabilities in the "Examine and Analyze Accounts

Data" stage (see Section 5.7), the information can be used to collect data from remote sources. This, however, requires the use of the `Transmit` and `Listen` capabilities, to communicate with external systems, and the associated constraints on the operation of these capabilities apply. Combining the encryption keys that we located with app data collected from the external storage, we were able to use the `Decrypt` capability to access files which had been protected by apps, presumably to prevent external viewing. Finally cached data presented a particularly rich valuable source within the internal storage, in terms of potential evidence data.

## 5.5 Examine App Files on External Storage

Again utilizing the `Forensic Examination` capability, we were able to locate files stored by the examined apps. The majority of these files were cached versions of files stored on the respective online services by the users and were stored unprotected. However, a subset of these files was stored using encryption, with the encryption key generally being stored within private storage.

## 5.6 Examine App Databases

The `Forensic Examination` of app databases resulted in the location of numerous artefacts of PII and file related metadata. PII data included user identifiers such as usernames and email addresses, access times, and geolocation data. The majority of data stored by the apps related to the files accessed and cached by them. File metadata included typical entries such as filename, paths, access, modification, deletion, synchronization and creation timestamps, and file types. The databases also stored a range of more esoteric metadata such as sharing, ownership and permissions metadata, encryption metadata, file hashes, JSON encoded data and file URLs. Forensic practitioners can combine select data, such as file URLs, with the `Listen` and `Transmit` capabilities to access data stored on remote systems. This may include files stored by the user or public information stored by the service about the user (such as avatars). However, consideration must be given to the constraints on these capabilities. For example, the forensic soundness requirement to ensure that *meaning* is preserved may be difficult to maintain when the precise purpose of a URL, located within an app's database, is unknown.

## 5.7 Examine and Analyze Accounts Data

Examination and analysis of accounts data utilizes a number of adversary capabilities, which if used without due consideration, may result in a breach of forensic soundness principles. For example, to extract accounts information, which is protected by Android's OS security, requires the use of the `Inject` capability. `Inject` is used, in concert with `Modify`, to add code to and modify the existing underlying OS frameworks, which ultimately results in the bypassing of security checking code (usually used to validate that the app which is requesting access to stored credentials is the same app that originally stored them). These functions would often need to be utilized on the evidence source device. This has the obvious consequence of potentially introducing *errors* and violating the associated forensic soundness requirements dealing with data modification. To avoid this circumstance, practitioners must ensure that the code being used is subject to *transparency and trustworthiness* constraints (i.e. having the code independently verified to ensure that it does not result in unknown modification to evidential data) and logging the operation of the code to ensure that the output is free of *errors*. *Experience* is another forensic soundness constraint that must be considered in the operation of these capabilities. A practitioner with *experience* in the development and operation of the `Injected` and `Modified` code is less likely to cause *errors* and ensure that *meaning* is maintained. After the data has been collected, the `Transmit` capability can be used to transfer the data from the mobile device to a forensic workstation for further examination and use.

While host-based intrusion detection systems (HIDS) [24] and other similar heuristic technologies are not yet prevalent on mobile devices, it is reasonable to assume that they may be introduced in the future as device performance improves [25]. With this in mind, the inclusion of a HIDS in a mobile device OS may increase the difficulty for a forensic practitioner when attempting modifications, such as this one, on the mobile device. This highlights the importance of live booting a forensic OS, which would bypass any HIDS implemented on the native OS. It should be possible to modify the OS frameworks on a forensic OS to allow for the extraction of credentials; however, this would significantly increase the level of per device customisation required for the forensic OS.

Examination of accounts data, in our case, returned a range of data including refresh tokens, access tokens, usernames, passwords, emails and timestamps. Refresh tokens are requested from the app servers using the users credentials and stored by apps to request updated access tokens which are, in turn, used to access protected resources (e.g. user files and records stored on remote systems).

### 5.8 Analyze App

After all available evidential data has been extracted from private and external storage, databases and OS accounts storage, a forensic practitioner may choose to conduct further analysis on the apps themselves. This analysis may be necessary to meet the *meaning* forensic soundness requirement as, although data may have been extracted, its precise meaning is not necessarily known. For example, timestamps may be recorded for "access" to an app or file, however it may not always be clear how "access" is defined by the app developers. Detailed analysis of the app would answer this question. This analysis was achieved in our case using both the decompilation of the app source code (static analysis) and analysis of the memory and operation of the app while it is running on a physical or emulated device (dynamic analysis). Using these techniques, we were able to locate various items of interest such as URLs for authentication, app secrets and protected information, which would otherwise be obfuscated during normal app operation.

### 5.9 Comparative Summary

Table 3 presents a comparison of this Android evidence collection and analysis methodology with similar methodologies in the literature.

As the work by and Chen, Yang and Liu [15] require the Android device to have an external SD card slot, they are unable to perform bit-for-bit physical copies of devices that do not allow for external SD cards. Furthermore, there may already be data on the obtained device's SD card. Their work cannot be applied to devices that do not have an SD card slot. Our methodology is applicable to devices with and without SD card slots and is able to perform bit-for-bit copies of both kinds of devices.

## Conclusion

Mobile devices and apps are an increasingly ubiquitous tool used throughout the daily lives of people worldwide. However these devices and apps can present a genuine security and privacy threat to mobile device and app users, due to their capacity to access sensitive data and personally identifiable information. In addition, as the use of mobile devices and apps grows, so too does their use by criminals, particularly in areas of sophisticated and organized crime where ongoing secure communications is critical for the operation of a criminal syndicate.

Table 3. A comparison of Android data collection methodologies.

| Requirements or Outcomes | Our Adversary Model Derived Methodology [5] (Fig 1) | Votipka, Vidas and Christin [16] | Vidas, Zhang and Christin [13] | Son, et al. [17] | Lessard and Kessler [14] | Chen, Yang and Liu [15] |
|---|---|---|---|---|---|---|
| Require device to be rooted? | **No** | No | No | No | Yes | No |
| Recovery/boot partition flashed | **No** | Yes, recovery partition | Yes, recovery partition | Yes, boot partition | No | Yes, recovery partition |
| Collects bit-for-bit physical copy? | **Yes** | Yes | Yes | No | No | No |
| Collects secure credential storage? | **Yes** | No | No | No | No | No |
| External SD Card required? | **No** | No | No | No | Yes | Yes |
| Analyzes collected data? | **Yes** | No | No | No | Yes | No |

doi:10.1371/journal.pone.0138449.t003

In this paper, we proposed a forensic adversary model which, to the best of our knowledge, is the first adversary model to be used in a forensic context. By utilizing the McKemmish [20] model for forensic soundness as a base in terms of selecting capabilities, the adversary model formalizes the real world capabilities of a forensic practitioner in obtaining evidential data from mobile devices. A key difference between our forensic adversary model and the adversary model(s) used in security and cryptography research is the constraint of forensic soundness (see Table 3).

We demonstrated how this adversary model can be used to derive an Android evidence collection and analysis methodology. This methodology facilitates the presentation of valid evidence. Using the methodology, we analyzed a number of cloud-based apps for potential evidential data. We then summarized the relationship between our findings, including the key forensic artefacts that we were able to locate, and the adversary's capabilities and forensic soundness constraints. We also found that it was possible to obtain a significant amount of evidential data using our methodology whilst also maintaining the required level of forensic soundness (outlined in Definition 1).

Future work includes deploying the forensic adversary model in a range of forensic case studies and investigations.

## Acknowledgments

The views and opinions expressed in this paper are those of the authors alone and not the organizations with whom the authors are or have been associated. The authors would also like to thank the anonymous reviewers for providing constructive and generous feedback. Despite their invaluable assistance, any errors remaining in this paper are solely attributed to the authors.

## Author Contributions